\documentstyle[prb,aps]{revtex}

\begin{document}
\author{A.V.Dooglav$^{1,2}$, H.Alloul$^{2}$, O.N.Bakharev$^{1,3}$, C.Berthier$^{3}$,
A.V.Egorov$^{1,4}$, M.Horvatic$^{3}$, E.V.Krjukov$^{1}$, P.Mendels$^{2}$,
Yu.A.Sakhratov$^{1}$, M.A.Teplov$^{1}$}
\address{$^{1}$Magnetic Resonance Laboratory, Kazan State University, 420008 Kazan,\\
Russia\\
$^{2}$Laboratoire de Physique des Solides, Universit\'{e} de Paris-Sud, 91405%
\\
Orsay Cedex, France\\
$^3$Grenoble High Magnetic Field Laboratory, Max-Plank Institut f\"ur\\
Festk\"orperforschung and Centre National de la Recherche Scientifique, BP\\
166, 38042 Grenoble Cedex 9, France\\
Laboratoire de Spectrom\'etrie Physique, Universit\'e Joseph Fourier,\\
Grenoble 1, 38402 Saint-Martin d'H\`eres Cedex, France\\
$^4$KFA Forschungszentrum J\"ulich, IFF, D-52425 J\"ulich, Germany}
\title{Cu(2) nuclear resonance evidence for an original magnetic phase in aged
60K-superconductors RBa$_{2}$Cu$_{3}$O$_{6+x}$ (R=Tm, Y)}
\maketitle

\begin{abstract}
\newline
It is widely believed that the long-range antiferromagnetic order in the RBa$%
_{2}$Cu$_{3}$O$_{6+x}$ compounds (R=Y and rare earths except of Ce, Pr, Tb)
is totally suppressed for the oxygen index $x\geq 0.4$ (AFM insulator-metal
transition). We present the results of the copper NQR/NMR studies of aged RBa%
$_{2}$Cu$_{3}$O$_{6+x}$ (R=Tm,Y) samples showing that a magnetic order can
still be present at oxygen contents $x$ up to at least 0.7 and at
temperatures as high as 77K.
\end{abstract}


\section{INTRODUCTION}

An enormous amount of papers, both experimental and theoretical, are devoted
to the NMR and NQR studies of copper in superconducting YBa$_{2}$Cu$_{3}$O$%
_{6+x}$ layered cuprates. It is considered well known that in
non-superconducting materials ($x<0.4$) nuclei of copper located in the CuO$%
_{2}$ planes experience the influence of a strong internal magnetic field $%
H_{int}\approx 80$ kOe which is perpendicular to the $c-$ axis, and that in
superconductors ($x>0.4$) such field is absent and $^{63,65}$Cu(2) nuclei
with the spin $I=3/2$ give typical NQR spectra at frequencies of $\nu
_{Q}\approx 25\div 32$ MHz, i.e., in the same frequency range as that for
NQR spectra of two-fold coordinated ''chain'' copper $^{63,65}$Cu(1) \cite
{ref1}. While studying NMR of thulium in oxygen-deficient TmBa$_{2}$Cu$_{3}$O%
$_{6+x}$ compounds \cite{ref2} we have discovered the wide and subtle
absorption line (Fig.1) which looked like copper NMR line but could not have
been attributed to the mentioned Cu(2) and Cu(1) centers with $\nu
_{Q}\approx 25\div 32$ MHz and $H_{int}=0$. We consequently undertook
experiments with TmBa$_{2}$Cu$_{3}$O$_{6+x}$ samples ($x=0.51$, 0.6, 0.7) in
zero external field at T=4.2K, which allowed us to observe a Cu(2) ZFNMR
spectrum completely different from those described in the literature. The
results of these experiments are presented in this paper. The analysis of
Cu(2) spectra of both types corresponding to $H_{int}\neq 0$ (type $I$, $\nu
_{ZFNMR}=55-135$ MHz) and $H_{int}=0$ (type $II$, $\nu _{Q}\approx 25\div 32$
MHz) shows that each of them belongs to two varieties of Cu(2) centers, $A$
and $B$, possessing different NQR frequencies and that $A_{I}-$ and $B_{I}-$%
centers differ much in the value of $H_{int}$. Similar results are obtained
for one YBa$_{2}$Cu$_{3}$O$_{6.66}$ sample\cite{ref3}.

\section{EXPERIMENTAL}

The TmBa$_{2}$Cu$_{3}$O$_{6+x}$ samples used in these experiments were $c$%
-axis-oriented powders mixed with paraffin (hereafter denoted as Tm6+x).
They were previously used for Tm NMR studies \cite{ref2}. All of them were
stored at room temperature for almost 6 years after preparation. Critical
temperatures (T$_{c}^{onset}$=53K, 61K and 64K for $x=0.51$, 0.6 and 0.7,
respectively) were obtained from measurements of the diamagnetic
susceptibility at a frequency of 1 kHz. Home-built spin-echo NQR
spectrometers were used for the Cu(2) NQR and ZFNMR measurements. Examples
of the copper NQR spectra in Tm6.6 are shown in Fig.2. Comparing the
non-saturated (NS) and saturated (S) spectra (Figs.2a,b) taken at 4.2K with
pulse sequence repetition rates of 1 and 200 Hz, respectively, one can
distinguish two contributions with different spin-lattice relaxation times $%
T_{1}$ (Fig.2c) and separate them (Figs.2d,e) by using a simple subtraction
procedure. Two narrow lines in Fig.2e corresponding to the long $T_{1}$ time
are believed to originate from the two-fold coordinated Cu(1)$_{2}$ sites
which belong to ''empty chains'' surrounded by ''full chains'' in the CuO$%
_{x}$ plane \cite{ref1}. The broad spectrum in Fig.2d characterized by the
short $T_{1}$ has a two-hump shape typical for the Cu(2) NQR spectrum of an
annealed sample \cite{ref4}. In the following it is denoted as the spectrum
of type $II$. Computer simulations have shown that this two-hump spectrum
can be described by a sum of four Gaussians (two sites, $A_{II}$ and $B_{II}$%
, times two isotopes, $^{63}$Cu and $^{65}$Cu), the amounts of $A_{II}$- and 
$B_{II}$-sites appearing in a ratio of approximately 2:1 independent of the
oxygen index $x$. The parameters of the spectra taken with a fixed pulse
separation time $\tau =25\mu s$ are given in Table I for three samples
studied.

The Cu(2) ZFNMR (type $I$) spectra taken at 4.2K with a fixed pulse
separation time $\tau =33\mu s$ are shown in Fig.3. A similar spectrum was
observed in Tm6.6 at a temperature of 77K (shown by filled squares in
Fig.3b). It should be noted here that in our experiments the well pronounced
spectra of type $I$ were observed only in those Tm- and Y-compounds which
exhibited well pronounced two-hump spectra of type $II$ (Fig.2d). An obvious
downfall of the spectral intensity at around 98 MHz which looks particularly
pronounced at $x$=0.7 clearly shows that the ZFNMR spectrum is also composed
of two contributions, those are the low-frequency fragment (LFF) at 55-98
MHz and the high-frequency fragment (HFF) at 98-135 MHz.

In order to estimate the integrated intensities, $S_{I}$ and $S_{II}$, of
the ZFNMR (type $I$) and NQR (type $II$) spectra, we have corrected them
according to the spin-spin relaxation times $T_{2}$ measured at the main
peaks of the spectra. All spin-echo decays measured at 4.2K have been found
to obey the following formula: $A_{2\tau }/A_{0}=exp(-(2\tau /T_{2})^{n})$.
In the series of Tm compounds, the longest $T_{2}$-times were measured for
the Tm6.6 sample: $T_{2}$=72, 64, 119, 75, 43 and 54 $\mu s$ at frequencies
of 62, 74, 87, 109, 116 and 124 MHz, respectively; the values of $n$ were
found to lie in the range 0.7-0.9. Because of the broad frequency range the
echo intensity has also been corrected by a factor $\nu ^{2}$, one factor $%
\upsilon $ due to the nuclear magnetization, the other one due to the
precession frequency of the nuclear magnetization. The $T_{2}$-correction
procedure consisted of four steps: (i) fitting the experimental Cu(2)
spectra by a superposition of Gaussians (6 for type $I$ and 4 for type $II$%
); (ii) multiplying intensity of each component by the corresponding factor $%
exp((2\tau /T_{2})^{n})$; (iii) summation of the corrected Gaussian lines;
(iv) $\nu ^{2}$ correction. The resulting spectra are shown by solid lines
in Fig.4. The results of the best fit for the Tm6.51 and Tm6.6 samples show
the LFF- and HFF-intensities to be in a ratio $\approx 2:1$ which allows us
to ascribe them, in analogy with the above $II$-type spectrum, to the $A_{I}$%
- and $B_{I}$-sites, respectively. In Table II, the partial ratios $%
S_{AI}/S_{BI}$ and $S_{AII}/S_{BII}$ for spectral intensities of the $A$-
and $B$-sites are given along with the ratio $S_{II}/S_{I}$. The absolute
values of the signal intensities per 1 gram of material, $S_{I}$ and $S_{II}$%
, shown in the last column of Table II, are normalized to the intensity of
the ''ordinary'' Cu(2) ZFNMR spectrum of the TmBa$_{2}$Cu$_{3}$O$_{6.1}$
antiferromagnet as observed at T=4.2K in the frequency range from 61 to 123
MHz.

As can be seen from the last column of Table II, the volume fraction of the
material contributing to the total Cu(2) resonance absorption varies from
sample to sample being equal to 1.00, 0.58 and 0.83 in Tm6.51, Tm6.6 and
Tm6.7, respectively, as compared to Tm6.1. Since the Cu(2) ZFNMR spectrum is
certainly produced by a non-superconductor, the question arises: What part
of the Cu(2) NQR signal intensity can be attributed to the superconducting
fraction? To answer this question, we have re-measured the ac diamagnetic
susceptibility of the Tm6+x samples (x=$0.51\div 1.0$)\cite{ref2} at a
temperature of 4.2 K. The results of the measurements (see inset in Fig.1)
clearly show that the superconducting fraction decreases with ageing time
(with no changes in T$_{c}$), the tendency being most pronounced for the
60K-superconductors, and that the superconducting fraction of the 6-years
old samples under study is quite small, i.e., 0.05, 0.05 and 0.15 for
Tm6.51, Tm6.6 and Tm6.7, respectively. Comparing the latter quantities with
the Cu(2) NQR signal intensities (0.79, 0.42 and 0.73, respectively), one
arrives at the conclusion that at least a part of the Cu(2) absorption
observed at 4.2K definitely originates from the non-superconducting
material. This conclusion is also supported by the fact that in one of the
samples studied, YBa$_{2}$Cu$_{3}$O$_{6.66}$, which was prepared quite
recently, we did not observe a two-hump NQR, nor a ZFNMR spectrum\cite{ref3}.

\section{DISCUSSION}

In order to clarify the origin of the LFF- and HFF-spectra, we have tried to
fit the Cu(2) ZFNMR in a correct way assuming that the $A_{I}$- and $B_{I}$%
-sites can be characterized by different values of $\nu _{Q}$, $H_{int}$ and 
$\theta $ (angle between {\bf H}$_{int}$ and the $c$-axis). Prior to fitting
the experimental spectra (Fig.3) have been $T_{2}$-corrected by the function 
$f(\nu )=F_{0}(\nu )/F_{2\tau }(\nu )$, where $F_{2\tau }(\nu )$ is the
ZFNMR spectrum shown in Fig.4 by a dotted line, and $F_{0}(\nu )$ is the
spectrum represented by a solid line. The spectra of Tm6.51 and Tm6.6
corrected in this way are shown in Fig.5. The fitting procedure included
numeric diagonalization of the two Hamiltonians and calculation of all the
transitions probabilities ($\sim \nu ^{2}$). When calculating averaged
values of the probabilities in oriented powdered samples, we have taken an
orientation of a radiofrequency field with respect to the $c$-axis into
account ({\bf H}$_{1}\perp c$). In all the cases considered, the Lorentzian
shape of an individual resonance line was found to fit the experimental data
much better than the Gaussian one. Two versions of the fit have been
considered. When the angles $\theta $ were allowed to be free, the best fit
of the Cu(2) ZFNMR spectrum of Tm6.51 was obtained at the following
parameters (see solid line in Fig.5a): 
\begin{equation}
\begin{array}{c}
\text{''}\theta \text{ free''} \\ 
A_{I}\text{-sites, }\nu =55-98\text{ MHz, intensity }S_{AI}=62(6)\%\text{, }
\\ 
\theta =82(4)^{\text{o}}\text{, }H_{int}=63.9(3)\text{ kOe, }\nu _{Q}=31(2)%
\text{ MHz,}
\end{array}
\label{eq1}
\end{equation}
\begin{equation}
\begin{array}{c}
B_{I}\text{-sites, }\nu =98-127\text{ MHz\cite{ref5}, intensity }%
S_{BI}=38(3)\%\text{,} \\ 
\theta =64(1)^{\text{o}}\text{, }H_{int}=97.7(2)\text{ kOe, }\nu _{Q}=25(2)%
\text{ MHz,} \\ 
\text{linewidth }FWHM=4.4(4)\text{ MHz.}
\end{array}
\label{eq2}
\end{equation}
The corresponding parameters for the Tm6.6 sample have been obtained as
follows (Fig.5b): 
\begin{equation}
\begin{array}{c}
A_{I}\text{-sites, }\nu =55-98\text{ MHz, }S_{AI}=66(4)\%\text{,} \\ 
\theta =83(3)^{\text{o}}\text{, }H_{int}=64.1(2)\text{ kOe, }\nu _{Q}=30(1)%
\text{ MHz,}
\end{array}
\label{eq3}
\end{equation}
\begin{equation}
\begin{array}{c}
B_{I}\text{-sites, }\nu =98-127\text{ MHz\cite{ref5}, }S_{BI}=34(2)\%\text{,}
\\ 
\theta =63.5(5)^{\text{o}}\text{, }H_{int}=97.6(1)\text{ kOe, }\nu _{Q}=27(1)%
\text{ MHz,} \\ 
FWHM=4.4(3)\text{ MHz.}
\end{array}
\label{eq4}
\end{equation}
In the second version ($\theta =90^{o}$) both internal fields were assumed
to lie in the $ab$ plane. The results of this fit for Tm6.6 are as follows
(see solid line in Fig.5c): 
\begin{equation}
\begin{array}{c}
\text{''}\theta =90^{o}\text{''*} \\ 
A_{I}\text{-sites, }\nu =55-98\text{ MHz, intensity }S_{AI}=66(6)\%\text{,}
\\ 
H_{int}=64\text{ kOe}^{\text{*}}\text{, }\nu _{Q}=30\text{ MHz}^{\text{*}}%
\text{,}
\end{array}
\label{eq5}
\end{equation}
\begin{equation}
\begin{array}{c}
B_{I}\text{-sites, }\nu =98-135\text{ MHz, }S_{BI}=34(3)\%\text{,} \\ 
H_{int}=103\text{ kOe}^{\text{*}}\text{, }\nu _{Q}=15.3(7)\text{ MHz,} \\ 
FWHM=4.8(4)\text{ MHz}
\end{array}
\label{eq6}
\end{equation}
($^{*}$ denotes the fixed parameters). The second fit seems to agree with
experiment worse than the first one, but this can be due to incorrect
intensities of the HFF-lines deduced from the T$_{2}$-correction procedure.
We believe that at the present stage both versions can be considered
satisfactory; more experimental work is needed to make a choice between
them. In both cases the Cu(2) NMR line in Fig.1 has been identified as
originating from two groups of $A_{I}$-lines (at 58-63 MHz and 74-79 MHz).

Then summarizing all experimental facts, one should adopt that at least a
part of the two-hump NQR spectrum, which is supposed to be typical for the
annealed 123 superconductors with high T$_c$s \cite{ref4}, and the ZFNMR
spectrum, which is observed in the 60K-superconductors exhibiting a well
pronounced two-hump NQR, are actually representing a non-superconducting
material. In what follows, we try to speculate about the plausible structure
of this material. It is known that a structural (chemical) micro-phase
separation takes place in the oxygen-deficient 123 superconductors \cite
{ref2,ref6,ref7,ref8,ref9}. Therefore, one could naturally expect to find
the defects of the crystal lattice in the form of layers (reflecting the
basic property of the layered structure itself) and, following this way,
could associate the two types of Cu(2) spectra with two different
non-superconducting micro-phases, i.e., hole doped (type $II$) and non-doped
(type $I$) layers, forming some kind of stacking sequence along the $c$-axis 
\cite{ref2}. On the other hand, the coincidence of the $\nu _Q$-values (see
Eqs.(1)-(4) and Table I) in the two absolutely different phases lead us to
consider that both types of Cu(2) spectra might originate from the same CuO$%
_2$ layers (bilayers).

To make a choice between these two scenarios, one should use the neutron
scattering data which gives a direct information on a magnetic structure of
the oxygen-deficient 123 compounds. The thorough studies of the neutron
scattering in YBa$_2$Cu$_3$O$_{6.6}$ \cite{ref10,ref11,ref12,ref13} have
resulted in a recent observation \cite{ref13} of incommensurate magnetic
fluctuations peaked at $\overrightarrow{Q}=(1/2\pm \delta ,1/2\pm \delta )$
with $\delta =0.057\pm 0.006$ r.l.u. It was also found that the dynamical
susceptibility $\chi ^{\prime \prime }(\overrightarrow{q},\omega )$ at the
incommensurate positions first appears at temperatures somewhat above T$_c$
and then increases on cooling below T$_c$, altogether with the suppression
of magnetic fluctuations at the commensurate points. These observations look
similar to those for the La$_{1.6-x}$Nd$_{0.4}$Sr$_x$CuO$_4$ compounds\cite
{ref14,ref15} and thus can be considered as giving evidence for a formation
of a static (or quasi-static) stripe pattern in the CuO$_2$ layers. The
layer-by-layer separation scenario seems to be inconsistent with such a
stripe model. On the contrary, the stripe-like modulation of charge and spin
densities in CuO$_2$ layers resulting in an inequivalency of Cu(2) sites
looks very likely. Indeed, analyzing the data of Table II, one can conclude
that there could exist a certain value of $x_{pin}$ (close to 0.6) which
corresponds to the following relations: 
\begin{equation}  \label{eq7}
S_{AI}/S_{BI}=2,\quad S_{AII}/S_{BII}=2,\quad S_{II}/S_I=2\text{.}
\end{equation}
These relations can be understood in the frame of the charge stripe model
suggested in \cite{ref16}. A plausible stripe pattern corresponding to the
optimally doped CuO$_2$ planes ($1/6$ hole per CuO$_2$ unit) is shown in
Fig.6a. If one takes every third stripe away (Fig.6b), the hole
concentration $p$ becomes $(2/3)\times (1/6)=1/9$. According to the
empirical formula $p=0.187-0.21(1-x)$, suggested by Tallon et al. \cite
{ref17} for x$\geq 0.45,$ the concentration $p=1/9$ corresponds to $x=$ 0.64
which is close to what is expected for Eq.(7) to hold. The magnetic
superstructure shown in Fig.6b has a period of 18 lattice spacings. This
period and the diagonal direction [110] of the stripes in Fig.6b exactly
corresponds to the incommensurate magnetic fluctuations observed in YBa$_2$Cu%
$_3$O$_{6.6}$ \cite{ref13}: the model in Fig.6b predicts incommensurate
peaks at $\overrightarrow{Q}=(1/2\pm \delta ,1/2\pm \delta )$ with $\delta
=1/18=0.0556$. In this stripe pattern, it is easy to distinguish four
different Cu(2) sites which could be identified as sites $A_I$ (at borders
of magnetic stripes), $B_I$ (at centers of magnetic stripes), $A_{II}$ (at
borders of non-magnetic stripes), $B_{II}$ (at centers of non-magnetic
stripes). An alternative possibility is to ascribe the $B_{II}$-sites to
copper atoms located at outer borders of the non-magnetic bi-stripes
allowing the $A_{II}$-sites to occupy four lines of coppers inside the
bi-stripes. In any case, however, one has a problem to explain why the
transferred hyperfine field from the $A_I$-copper spins does not influence
the NQR spectrum of the neighboring type-$II$ copper centers. Further
experimental work is necessary to better understand the allocation of the
Cu(2) centers responsible for the two-hump NQR spectrum.

It is known that the observation of the static stripe-phase order of holes
and spins in La$_{1.6-x}$Nd$_{0.4}$Sr$_x$CuO$_4$ \cite{ref14,ref15} has
appeared possible due to pinning of the stripes by the neodymium impurities.
What could be the reason for stripes to be pinned in non-stoichiometric 123
compounds? We believe that the so-called ''OrthoIII'' superstructure in CuO$%
_x$ planes could cause pinning of charge stripes in CuO$_2$ planes since
this superstructure has a period ($3a_0$) commensurate with the period of
the stripe pattern shown in Fig.6b. A possibility for the OrtoIII to form a
stable phase at $x=0.65-0.77$ was proved by the electron \cite{ref18}, X-ray
and neutron \cite{ref19} diffraction measurements. The above oxygen contents
appear to be shifted to higher values than that ($x_{pin}$) expected for the
condition Eq.(7) to be fulfilled. However, the value of $x_{pin}$ may happen
to result from an interplay between the OrthoIII ordering of CuO$_x$ layers
at $x>x_{pin}$ and an appropriate hole doping of CuO$_2$ layers ($p=1/9$ per
CuO$_2$ at $x<x_{pin}$). Experiments with an YBa$_2$Cu$_3$O$_{6.77}$ single
crystal \cite{ref19} have shown that the OrthoIII phase is stable at
temperatures below 75$^o$C. Therefore, the interactions responsible for the
formation of the OrthoIII phase seem to be rather weak and can be easily
disturbed by thermally-activated oxygen diffusion. This feature of the T-$x$
phase diagram allows us to understand why the Cu(2) ZFNMR was observed in
our experiments only in those 123 superconductors which were subjected to a
very long-term room-temperature annealing.

\section{CONCLUSIONS}

The Cu(2) nuclear resonance spectra were studied at liquid helium
temperatures in samples of oxygen-deficient 60K-superconductors, TmBa$_{2}$Cu%
$_{3}$O$_{6+x}$ and YBa$_{2}$Cu$_{3}$O$_{6.66}$, stored at room temperature
for a long time (up to 6 years). Cu(2) ZFNMR spectra different from those
known so far were observed, indicating the presence of a non-superconducting
phase in the superconducting samples. The quantitative analysis of the
copper resonance absorption intensities in different samples lead us to
consider that both types of Cu(2) spectra (ZFNMR at 55-135 MHz and at least
a part of NQR at 25-32 MHz) might originate from the same
non-superconducting CuO$_{2}$ layers decorated by the pinned charge stripes.

\section{ACKNOWLEDGEMENTS}

This work was supported in part by the Russian Scientific Council on
Superconductivity, under Project 94029, by the Russian Foundation for Basic
Research, under Project 96-02-17058a, by the NATO Scientific and
Environmental Affairs Division, under Grant HTEC.LG-950536, and by the
INTAS, under Grant 96-0393.

\newpage

\section{Figure captions}

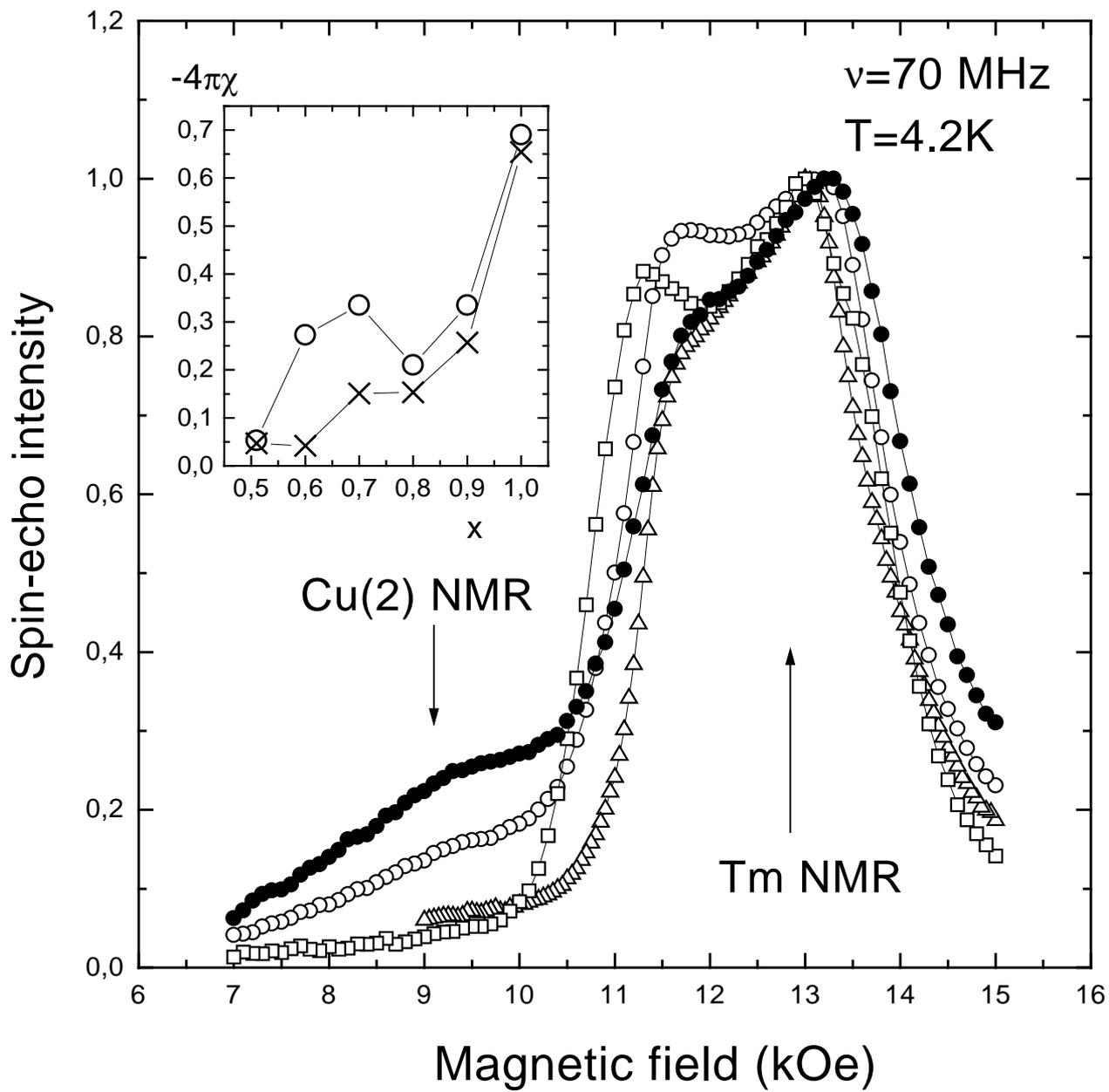
Fig.1. Tm and Cu(2) NMR spectra of the aligned Tm6+x powders with $x$=0.51
(triangles), 0.6 (light circles), 0.7 (squares) at a frequency of 70 MHz in
an external field H perpendicular to the $c$-axis. Cu(2) NMR is particularly
pronounced in the saturated spectrum of Tm6.6 (black circles) taken at a
pulse sequence repetition rate of F=100 Hz. Inset: ac diamagnetic
susceptibility of Tm6+x at liquid helium temperature (frequency of 1 kHz, H$%
_1\approx 1$ Oe); circles - for 1-year aged samples\cite{ref2}, crosses -
for 6-years aged samples.

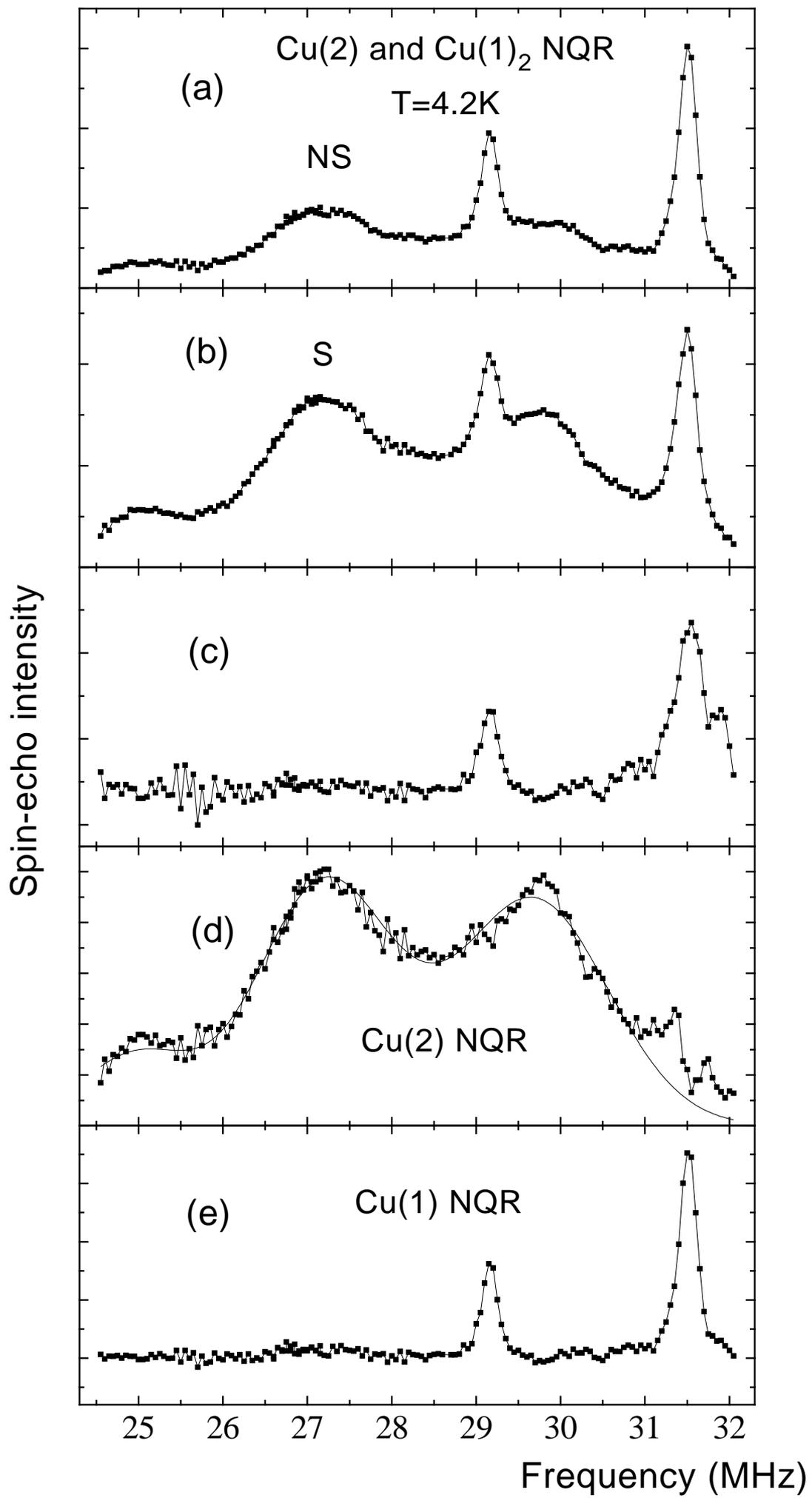
Fig.2. Copper NQR spectra of Tm6.6 at a temperature of 4.2K: (a)
non-saturated Cu(1) and Cu(2) spectra (F=1 Hz), (b) saturated Cu(1) and
Cu(2) spectra (F=200 Hz), (c) ratio of the non-saturated to saturated echo
intensity, (d) Cu(2) NQR spectrum, (e) Cu(1) NQR spectrum. Solid line in (d)
is an approximation by four Gaussians (for parameters see Table I).

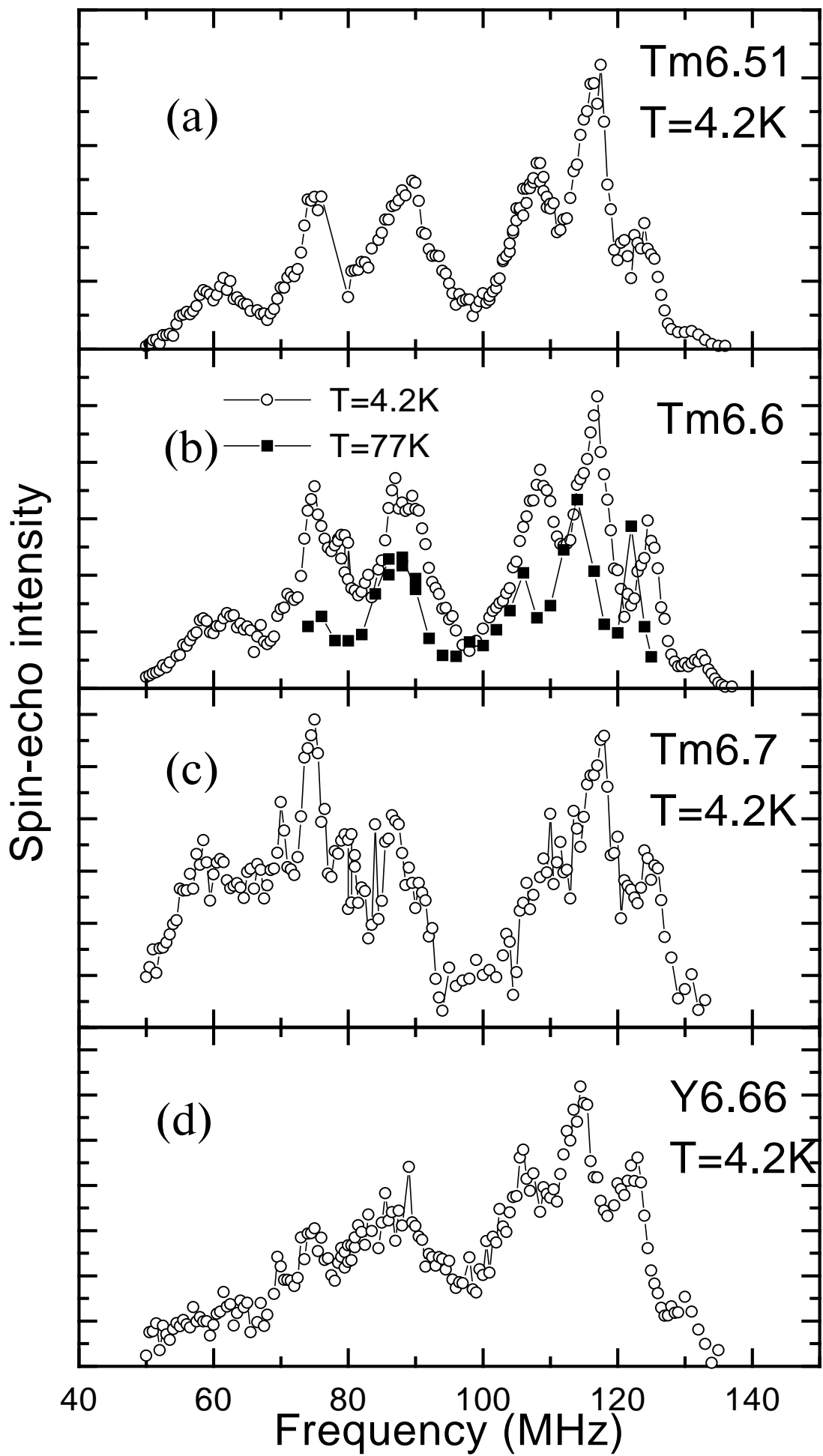
Fig.3. Cu(2) ZFNMR spectra of (a) Tm6.51, (b) Tm6.6, (c) Tm6.7 and (d) Y6.66
at a temperature of 4.2K. Black squares in (b) is the spectrum at T=77K
multiplied by a factor of 10. The YBa$_2$Cu$_3$O$_{6.66}$ sample was stored
at room temperature for 3 years after preparation.

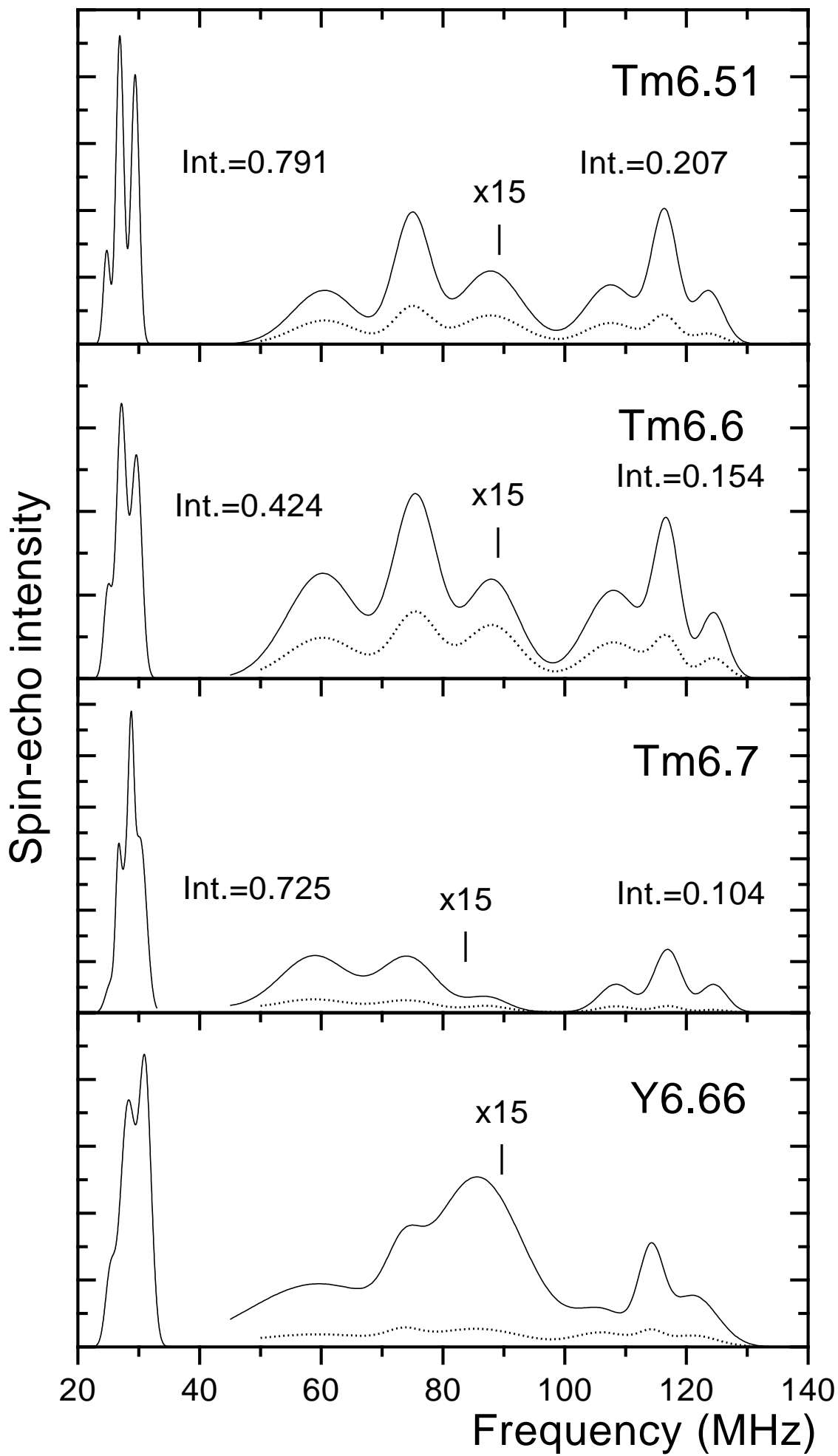
Fig.4. Calculated Cu(2) NQR and ZFNMR spectra of (a) Tm6.51, (b) Tm6.6, (c)
Tm6.7 and (d) Y6.66 obtained from the experimental spectra of Fig.3 as a
superposition of six Gaussians: solid lines F$_0$($\nu $) - taking into
account the $\nu ^2$-type dependence of the spin-echo intensities and the
losses of signal intensities due to spin-spin relaxation, dotted lines F$%
_{2\tau }$($\nu $) - taking into account the $\nu ^2$-dependence only (for
details see text). The integrated intensities of the $S_I$- and $S_{II}$%
-spectra depicted by solid lines are normalized by the intensity of the
Cu(2) ZFNMR spectrum of the 1 gram TmBa$_2$Cu$_3$O$_{6.1}$ sample at T=4.2K.
For convenience of comparison, the ZFNMR spin-echo intensities are
multiplied by a factor of 15.

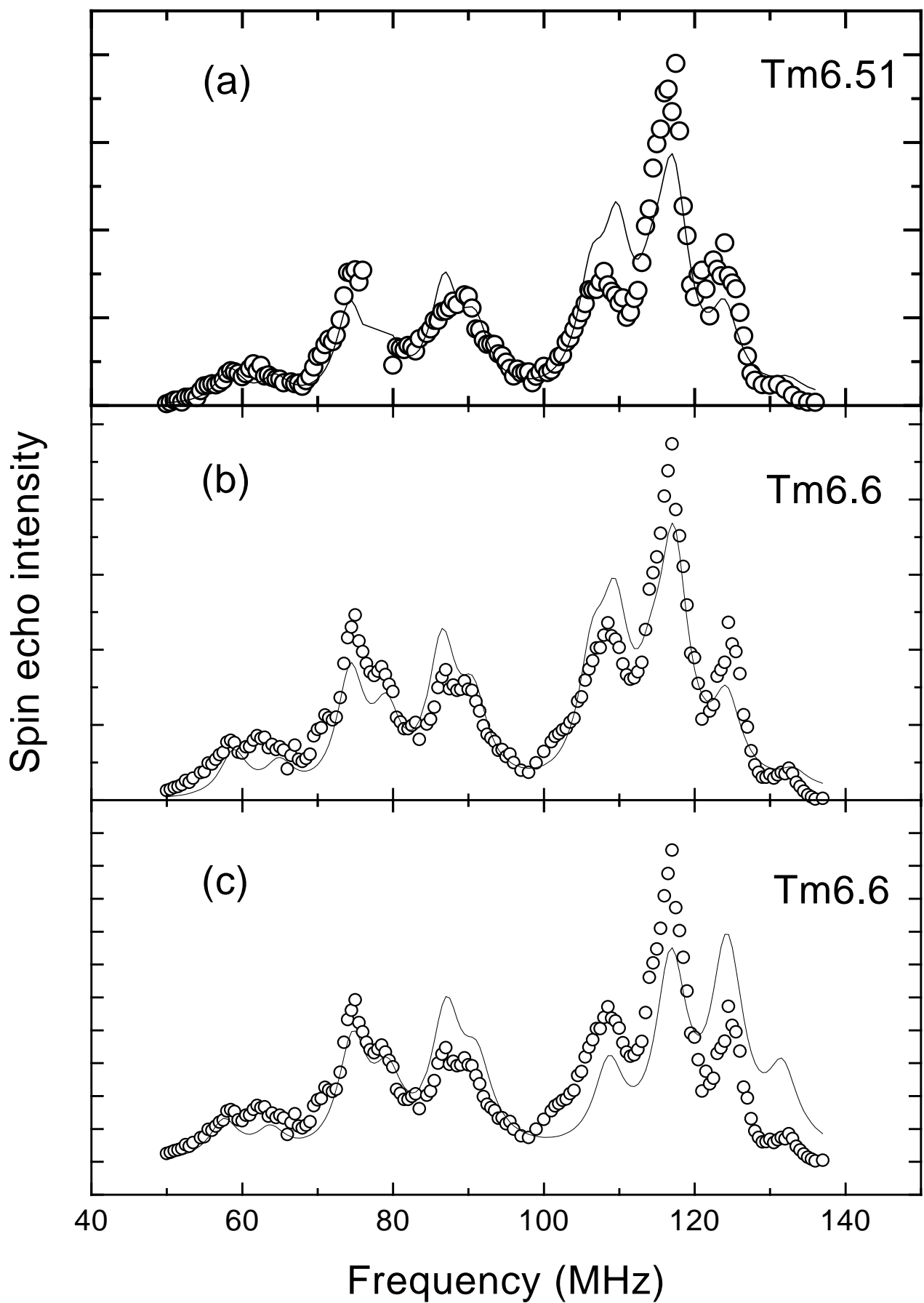
Fig.5. Cu(2) ZFNMR spectra of (a) Tm6.51 and (b) Tm6.6 obtained from the
experimental spectra of Figs.3a,b by taking into account the losses of
signal intensities due to spin-spin relaxation (for details see text). Solid
lines in (a) and (b) represent the results of the numerical diagonalization
of the two Hamiltonians with the parameters given in Eqs.(1),(2),(3),(4);
solid line in (c) represents the spectrum of Tm6.6 calculated with the
parameters given in Eqs.(5), (6).

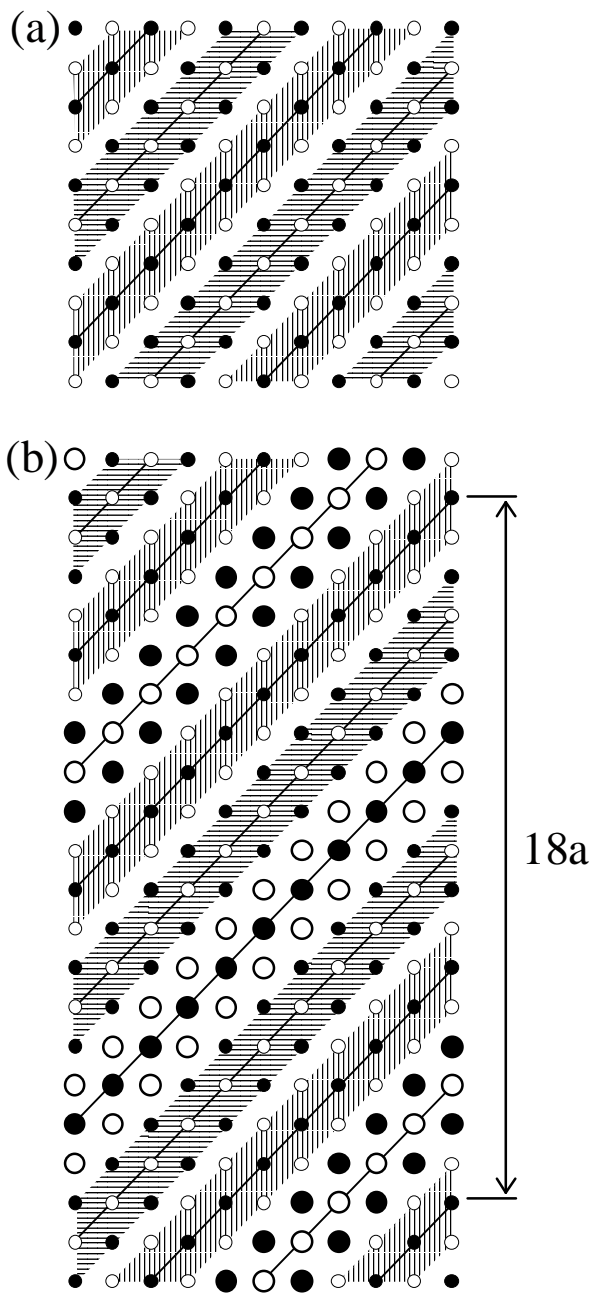
Fig.6. Plausible stripe patterns in the (a) optimally doped ($p=1/6$ hole
per Cu(2)) and (b) underdoped ($p=1/9$) CuO$_2$ planes. Copper atoms
carrying magnetic moments of opposite orientations are shown by big light
and black circles, those carrying no magnetic moments are shown by small
circles, solid lines mark the centers of the non-magnetic (hatched) and
magnetic (non-hatched) stripes. The period of 18 lattice spacings of the
magnetic superstructure in (b) is twice as large as the period of the charge
pattern.

\newpage

\mediumtext

\begin{table}[tbp]
\begin{tabular}{|c|c|c|c|c|}
\hline
&  &  &  &  \\ 
Sample & Site & $^{63}\nu_Q$ (MHz) & Full rms width (MHz) & Intensity \\ 
&  &  &  &  \\ \hline
Tm6.51 & $A_{II}$ & 29.44(2) & 1.35(3) & 0.65(1) \\ 
& $B_{II}$ & 26.76(6) & 1.17(4) & 0.35(4) \\ \hline
Tm6.6 & $A_{II}$ & 29.69(2) & 1.75(4) & 0.69(1) \\ 
& $B_{II}$ & 27.01(6) & 1.46(4) & 0.31(4) \\ \hline
Tm6.7 & $A_{II}$ & 30.36(3) & 2.02(6) & 0.68(1) \\ 
& $B_{II}$ & 27.39(6) & 1.79(9) & 0.32(2)
\end{tabular}
\caption{Parameters of the $^{63}$Cu(2) NQR spectra (type $II$) in Tm6+x
samples at T=4.2K (with no corrections for the frequency dependence and for
the spin-spin relaxation time $T_2$)}
\label{t1}
\end{table}

\begin{table}[tbp]
\begin{tabular}{|c|c|c|c|c|c|}
\hline
& \multicolumn{2}{c|}{ZFNMR} & \multicolumn{2}{c|}{NQR} & NQR/ZFNMR \\ 
Sample & \multicolumn{2}{c|}{$S_{AI}/S_{BI}$} & \multicolumn{2}{c|}{$%
S_{AII}/S_{BII}$} & $S_{II}/S_I$ \\ \cline{2-6}
& Lorentzian & Gaussian & Gaussian & Gaussian & Gaussian \\ 
& approximation & approximation & approximation & approximation & 
approximation \\ 
& with the $T_2$ & with the $T_2$ & with no $T_2$ & with the $T_2$ & with
the $T_2$ \\ 
& corrections & corrections & corrections & corrections & corrections \\ 
\hline
Tm6.51 & 1.6(3) & 1.6(4) & 1.8(2) & 1.8(6) & 0.791/0.207=3.8(1.3) \\ \hline
Tm6.6 & 2.0(2) & 2.0(5) & 2.2(2) & 2.0(7) & 0.424/0.154=2.8(9) \\ \hline
Tm6.7 &  & 3.7(1.2) & 2.1(2) & 2.5(9) & 0.725/0.104=7(2) \\ \hline
Y6.66 &  & 3.3(1.2) & 2.6(9) & 3.8(1.3) & 3.7(1.3)
\end{tabular}
\caption{Intensities of the Cu(2) spectra in Tm6+x (x=0.51, 0.6, 0.7) and
Y6.66 at T=4.2K}
\label{t2}
\end{table}


\begin{references}
\bibitem{ref5}  In the version ''$\theta $ free'' the Cu(2) ZFNMR line at a
frequency of 132 MHz (see Fig.3) was assumed to be a forbidden transition of
the $A_{I}$-sites.
\end{references}
\end{document}